\documentclass[aps,prd,twocolumn,groupedaddress,showpacs,preprintnumbers]{revtex4}

\usepackage{graphicx}
\usepackage{color}
\usepackage[T1]{fontenc}
\usepackage{amssymb,amsmath}
\usepackage{ulem}

\begin{document}


\title{Corrected constraints on big bang nucleosynthesis in a modified gravity model of $f(R) \propto R^n$}


\author{Motohiko Kusakabe$^{1,2}$}
\email{motohiko@kau.ac.kr} 
\author{Seoktae Koh$^{3}$}
\author{K. S. Kim$^{1}$}
\author{Myung-Ki Cheoun$^{2}$}
\affiliation{
$^1$School of Liberal Arts and Science, Korea Aerospace University, Goyang 412-791, Korea}
\affiliation{
$^2$Department of Physics, Soongsil University, Seoul 156-743, Korea}
\affiliation{
$^3$Department of Science Education, Jeju National University, Jeju 690-756, Korea}


\date{\today}

\begin{abstract}

Big bang nucleosynthesis in a modified gravity model of $f(R)\propto R^n$ is investigated.  The only free parameter of the model is a power-law index $n$.    We find cosmological solutions in a parameter region of $1< n \leq (4+\sqrt{6})/5$.  We calculate abundances of $^4$He, D, $^3$He, $^7$Li, and $^6$Li during big bang nucleosynthesis.  We compare the results with the latest observational data.  It is then found that the power-law index is constrained to be $( n-1 ) = ( -0.86 \pm 1.19 ) \times 10^{-4}$ (95 \% C.L.) mainly from observations of deuterium abundance as well as $^4$He abundance.
\end{abstract}


\preprint{Published in Phys. Rev. D {\bf 91}, 104023 (2015):  http://journals.aps.org/prd/abstract/10.1103/PhysRevD.91.104023.}
\preprint{Copyright 2015 American Physical Society}

\maketitle


\section{Introduction}\label{sec1}
The present standard cosmological model is based on Einstein's general relativity with the Friedmann-Lema\^{i}tre-Robertson-Walker metric for a homogeneous and isotropic universe.  All elementary particles of the standard particle model and dark matter and dark energy are taken into account in the cosmological model.  The standard cosmological model has been supported by various kinds of astronomical observations.  Observations of light element abundances in old astronomical objects are, however, one of the most important premises of the standard cosmological model.  Roughly speaking, the theoretical predictions of light element abundances are consistent with observational data.  In modified gravitational theories, cosmic expansion histories are different from that in the standard model, while in modified particle theories additional effects of exotic particles operate in the early universe.  As a result, primordial elemental abundances in these models are different from those in the standard big bang nucleosynthesis (BBN) model.  Therefore, we can limit any models which predict changes in abundances.

The baryogenesis in a modified gravity model of $f(R) \propto R^n$, where $R$ is the Ricci scalar and $n$ is the power-law index, has been studied to explain the small baryon-to-photon number ratio of the Universe \cite{Lambiase:2006dq}.  The authors derived a cosmological solution in which the scale factor of the universe scales as $a(t) \propto t^\alpha$, where $t$ is the cosmic time and $\alpha$ is a real parameter.  They argued that $(4-\sqrt{6})/5 \leq n \leq 1$ should be satisfied in order to realize a positive temperature of the universe.  The BBN in the same model has also been analytically studied \cite{Kang:2008zi}.  They constrained the index to be $1-n \lesssim 2 \times 10^{-4}$ by a comparison of an analytical estimation of $^4$He abundance and observational data.

In this paper, we calculate BBN in the model of $f(R) \propto R^n$ with a detailed nuclear reaction network code and show abundances of all light elements produced during BBN.  In Ref. \cite{Kang:2008zi}, only $^4$He abundance has been studied semianalytically.  In this paper, however, it is found that observational constraints on the primordial D abundance can limit the modified gravity model more stringently than those on the $^4$He abundance.  On the other hand, limits derived from observations of $^3$He, $^7$Li, and $^6$Li abundances are less stringent than those of D and $^4$He.  In addition, we point out that models of $f(R)$ should describe the accelerated expansion of the present Universe.  We find that the model used in the previous study \cite{Lambiase:2006dq,Kang:2008zi} is excluded by this requirement, and we suggest a simple correction to the model.  In this paper, we consider three models:  (1) a new model which describes the accelerated expansion of the present Universe, (2) the previous model \cite{Lambiase:2006dq,Kang:2008zi} which cannot describe the expansion, and (3) a corrected version of (2) which describes the expansion.  Although the limit on the $f(R) \propto R^n$ model is corrected, our revised result supports the previous conclusion that the consideration of BBN excludes parameter values of $n$ largely different from unity \cite{Kang:2008zi}.

In Sec. \ref{sec2}, the modified gravity model is introduced, and equations for the cosmic evolution are derived.  In Sec. \ref{sec3}, our code for the BBN calculation is briefly explained.  In Sec. \ref{sec4}, observational constraints on the primordial light element abundances are described.  In Sec. \ref{sec5}, a result of BBN is shown and interpreted.  In Sec. \ref{sec6}, this work is briefly summarized.

\section{Cosmology of $f(R)\propto R^n$ gravity}\label{sec2}
In this section, formulas of the cosmology in the modified gravity model are shown.  First, we derive equations of motion.  The action is given by
\begin{equation}
\label{eq1}
S= \frac{1}{2 \kappa^2} \int d^4 x \sqrt{-g} f(R) +S_\mathrm{m} (g_{\mu \nu}, \phi_\mathrm{m} ),
\end{equation}
where
$\kappa^2 =8\pi G$ is defined, with $G$ Newton's constant,
$g_{\mu \nu}$ the metric tensor,
$g$ the determinant of the metric tensor, and
$S_\mathrm{m}$ the action of the matter field $\phi_\mathrm{m}$ which takes into account radiation and matter in the Universe.
The field equation for gravity is then derived by varying this action with respect to the metric tensor,
\begin{equation}
\label{eq2}
f' R_{\mu \nu} -\frac{1}{2} f g_{\mu \nu} - \nabla_\mu \nabla_\nu f' +g_{\mu \nu} \Box f' = \kappa^2 T_{\mu \nu},
\end{equation}
where
$f' =df/dR$ is defined, and
$T_{\mu \nu}$ is the energy-momentum tensor for matter defined as
\begin{equation}
\label{eq_a11}
T_{\mu \nu} = - \frac{2}{\sqrt{-g}} \frac{\delta \left( \sqrt{ -g} {\cal L}_\mathrm{m} \right)}{\delta g^{\mu \nu}}.
\end{equation}
Here
${\cal L}_\mathrm{m}$ is the Lagrangian density of matter, and it is related to $S_\mathrm{m}$ by
\begin{equation}
\label{eq_a12}
S_\mathrm{m} = \int d^4 x \sqrt{-g} {\cal L}_\mathrm{m}.
\end{equation}
We assume the spatially flat Friedmann-Lema\^{i}tre-Robertson-Walker metric, as supposed in the standard cosmological model,
\begin{equation}
\label{eq3}
ds^2 = dt^2 -a(t)^2 \left( dx^2 +dy^2 +dz^2 \right).
\end{equation}
For matter, on the other hand, we assume a perfect fluid described by a time-dependent energy density $\rho(t)$ and pressure $p(t)$,
\begin{equation}
\label{eq4}
{T^\mu}_\nu = \mathrm{diag}\left( \rho, -p, -p, -p \right).
\end{equation}
The 0-0 component of Eq. (\ref{eq2}) then becomes 
\begin{equation}
\label{eq5}
-3 \frac{\ddot{a}}{a} f' -\frac{1}{2} f +3 \frac{\dot{a}}{a} f'' \dot{R} = \kappa^2 \rho.
\end{equation}
The $i$-$i$ components, on the other hand, give 
\begin{equation}
\label{eq6}
\left( \frac{\ddot{a}}{a} +2 \frac{\dot{a}^2}{a^2} \right) f'  +\frac{1}{2} f -2 \frac{\dot{a}}{a} f'' \dot{R} -f''' \dot{R}^2 -f'' \ddot{R} = \kappa^2 p.
\end{equation}
From Eqs. (\ref{eq5}) and (\ref{eq6}), the following energy conservation holds,
\begin{equation}
\label{eq7}
\dot{\rho} +3 \frac{\dot{a}}{a} \left( \rho +p \right) =0.
\end{equation}

Second, we constrain the model space to be studied in this paper.  We assume {two functional shapes for $f(R)$.  The first model is given \cite{Lambiase:2006dq,Kang:2008zi} by 
\begin{equation}
\label{eq_a2}
f_1(R) = \left( \frac{R}{A} \right)^n,
\end{equation}
where
$A$ is a constant given by $A = M_\mathrm{p}^{2-2/n}$, with $M_\mathrm{p}=1.22 \times 10^{19}$ GeV the Planck mass.  The power-law index $n$ is the only free parameter, and $n=1$ reduces to Einstein's general relativity.  
This model has been analyzed in Refs. \cite{Lambiase:2006dq,Kang:2008zi}, and proper solutions of the scale factor exist only for $(4-\sqrt{6})/10 \leq \alpha \leq 1/2$ in this model.

The second model is given \footnote{When another choice of the signs in the metric, $g_{00}=-1$ and $g_{ij}=a(t)^2 \delta_{ij}$, is adopted, the formulation is somewhat different from the present case.  The equations of motion are different from Eqs. (\ref{eq5}) and (\ref{eq6}), and the Ricci scalar, denoted as $R^\ast$, is opposite in sign to $R$ [Eq. (\ref{eq_a1})].  We find that, in general, the different choices of the sign in the metric result in different equations for the evolution of $\rho$ and $p$ [cf. Eqs. (\ref{eq10}) and (\ref{eq11})] if the same function, i.e., $f(R)$ and $f(R^\ast)$, respectively, is used.  The functions $f_1(R^\ast)=-(-R^\ast/A)^n$ and $f_2(R^\ast)=(R^\ast/A)^n$ then correspond to $f_1(R)$ and $f_2(R)$, respectively, in the present metric case.  In the case of $n=1$, however, it follows that $f_1(R)=f_2(R)=R/A$, and we obtain the same equations for $\rho$ and $p$ from Eq. (\ref{eq2}) independently of the choice of metric.}
by
\begin{equation}
\label{eq8}
f_2(R) = -\left( \frac{-R}{A} \right)^n.
\end{equation}
A formulation of this model is given using the same assumptions as those adopted in Refs. \cite{Lambiase:2006dq,Kang:2008zi}, as follows.
It is} assumed that the matter part is predominantly contributed by the radiation with $p=\rho/3$.  In this case, Eq. (\ref{eq7}) leads to a relation of $\rho \propto a^{-4}$.  Here, we additionally constrain the model space by assuming the power-law solution of the scale factor, i.e.,
\begin{equation}
\label{eq9}
a(t) \propto t^\alpha.
\end{equation}
Inserting Eqs. (\ref{eq8}) and (\ref{eq9}) into Eqs. (\ref{eq5}) and (\ref{eq6}), we have two equations,
\begin{equation}
\label{eq10}
- \frac{n\left(2n + \alpha -3 \right) }{2 \left( 2\alpha -1 \right) } +\frac{1}{2} = \kappa^2 \rho \left[ \frac{6 \alpha \left( 2\alpha -1 \right)}{At^2} \right]^{-n}
\end{equation}
and
\begin{eqnarray}
\label{eq11}
&&\frac{n\left(3\alpha -1 \right) }{6 \left( 2\alpha -1 \right) } +\frac{n \left( n-1 \right) \left(2\alpha -2n +1 \right) }{3 \alpha  \left( 2\alpha -1 \right) } -\frac{1}{2} \nonumber \\
&&~~~~~~~~~~= \kappa^2 p \left[ \frac{6 \alpha \left( 2\alpha -1 \right)}{At^2} \right]^{-n}.
\end{eqnarray}
We note that the Ricci scalar is given by
\begin{equation}
\label{eq_a1}
R = -6 \left[ \frac{\ddot{a}}{a} + \left( \frac{\dot{a}}{a} \right)^2 \right] 
= -\frac{6 \alpha \left( 2\alpha -1 \right)}{t^2}.
\end{equation}
It is found that $\alpha =n/2$ must be satisfied in order to hold Eqs. (\ref{eq10}) and (\ref{eq11}) for any time $t$.  Then we assume $\alpha =n/2$ in what follows.  In this case, the energy density is related to the cosmic time as
\begin{equation}
\label{eq12}
\rho = \frac{1}{\kappa^2} \left[ \frac{6 \alpha \left( 2\alpha -1 \right)}{At^2} \right]^{n} \frac{-10 \alpha^2 +8\alpha -1}{2 \left( 2\alpha -1 \right)}.
\end{equation}
The pressure satisfies the relation $p=\rho/3$.  Then we utilize the relation between the energy density and the cosmic temperature,
\begin{equation}
\label{eq13}
\rho = \frac{\pi^2}{30} g_\ast T^4,
\end{equation}
where
$T$ is the temperature and
$g_\ast(T)$ is the relativistic degrees of freedom for energy density.  From Eqs. (\ref{eq12}) and (\ref{eq13}), the time-temperature relation of the universe is derived as
\begin{equation}
\label{eq14}
T = \left( \frac{15}{4 \pi^3 g_\ast } \right)^{1/4} g_{2\alpha}^{1/4} \frac{M_\mathrm{p}^{1/2}} {t^\alpha A^{\alpha/2}},
\end{equation}
where
\begin{equation}
\label{eq15}
g_{2\alpha} = \left( 6 \alpha \right)^{2 \alpha} \frac{-10 \alpha^2 +8\alpha -1}{2 \left( 2\alpha -1 \right)^{1-2\alpha}}
\end{equation}
is defined.  Since the cosmic temperature must be positive, the parameter $g_{2\alpha}$ must be positive.  Then a constraint on $\alpha$ is derived,
\begin{equation}
\label{eq16}
\frac{1}{2} \leq \alpha \leq \frac{4+\sqrt{6}}{10}.
\end{equation}
The Hubble expansion rate is given as a function of energy density by inserting Eq. (\ref{eq12}) into the equation $H=\dot{a}/a =\alpha/t$.  We thus obtain
\begin{equation}
\label{eq17}
H = \alpha g_{2\alpha}^{-1/\left( 4\alpha \right)} M_\mathrm{p} \left( \frac{8\pi \rho} {M_\mathrm{p}^4} \right)^{1/ \left( 4\alpha \right)}.
\end{equation}
Finally, the Hubble expansion rate can also be directly given as a function of temperature using Eq. (\ref{eq13})
\begin{equation}
\label{eq20}
H = \frac{\alpha A^{1/2}}{g_{2\alpha}^{1/\left(4 \alpha\right)} M_\mathrm{p} ^{1/\left( 2\alpha \right) }}\left( \frac{4 \pi^3 g_\ast }{15} \right)^{1/\left( 4 \alpha \right)} T^{1/\alpha}.
\end{equation}
The special case of $n=1$ ($\alpha=1/2$) corresponds to Einstein's general relativity.
We note that the allowed region [Eq. (\ref{eq16})] is outside the region,  $(4-\sqrt{6})/10 \leq \alpha \leq 1/2$, in the $f_1(R)$ model \cite{Lambiase:2006dq,Kang:2008zi}.

The two models [Eqs. (\ref{eq_a2}) and (\ref{eq8})] successfully describe solutions of $R \geq 0$ and $R \leq 0$, respectively.  In the present setup, the left-hand sides of Eqs. (\ref{eq5}) and (\ref{eq6}) are proportional to the function $f(R)$.  Since $f_1(R) \propto R^n$ should be real for real number $n$, the scalar $R$ must be non-negative.  This requirement on $R$ is important when we derive the constraint [Eq. (\ref{eq16})].  For the case of $f_2(R)$, on the other hand, the scalar $R$ must be non-positive.  Thus, in this model the power-law functions $f_1(R)$ and $f_2(R)$ must be real.  Since the model of $f_1(R)$ \cite{Kang:2008zi} also satisfies this requirement during the BBN epoch, it looks like a possible cosmological model.

The model $f_1(R)$ is, however, excluded since it cannot describe the late Universe, i.e., the $\Lambda$CDM model, whose energy density is dominated by the dark energy ($\Lambda$) and cold dark matter (CDM).  Astronomical observations indicate that the Ricci scalar becomes negative in the late Universe, which can be described with the CDM and dark-energy-dominated universe.  For example, in the standard cosmological model, the Ricci scalar [Eq. (\ref{eq_a1})] is given by
\begin{eqnarray}
\label{eq_a3}
-R &=&\kappa^2 \left( \rho -3p \right) + 4 \Lambda \nonumber \\
&=&
\left\{ \begin{array}{ll}
0 & ({\rm radiation~dominated~epoch}) \\
\kappa^2 \rho & ({\rm matter~dominated~epoch}) \\
4 \Lambda & (\Lambda~{\rm dominated~epoch}) \\
\end{array}
\right.
\end{eqnarray}
Since the present Universe is explained by a negative $R$ value, the negative $R$ should be consistently accommodated in the cosmological model.  If the $R$ value during BBN is positive as in the case of $f_1(R)$, a transition from $R>0$ to $R<0$ must occur in the Universe between the BBN and the present epochs.  While the $f_2(R)$ model can describe this late Universe, the $f_1(R)$ model cannot because of nonreal $f_1(R)$ values for $R<0$ and $n\neq 1$.  The latter model is therefore excluded.

The model $f_1(R)$ can, however, be corrected so that it is consistent with observational evidence of accelerated expansion of the present Universe.  One simple correction is given by
\begin{equation}
\label{eq_a4}
f(R) = \operatorname{sgn} (R) \left| \frac{R}{A} \right|^n.
\end{equation}
This function reduces to $f_1(R)$ [Eq. (\ref{eq_a2})] for $R\geq 0$ and $f_2(R)$ [Eq. (\ref{eq8})] for $R < 0$.  This function also describes the general relativity ($n=1$) in which $f(R)=R/A$ is satisfied.  It includes both solutions for $\alpha \geq 1/2$ and $\alpha < 1/2$, and allows a transition from $R>0$ to $R<0$ in the cosmic evolution.  Therefore, the model $f(R)$ is consistent with the acceleration of the Universe.  We then use this function in the following calculation.  When this model is used, the time-temperature relation is given by Eq. (\ref{eq14}) with the function $g_{2\alpha}$ replaced by
\begin{equation}
\label{eq_a5}
g_\alpha = \left( 6 \alpha \right)^{2 \alpha} \frac{-10 \alpha^2 +8\alpha -1}{2 \left| 2\alpha -1 \right|^{1-2\alpha}}.
\end{equation}

We show that for any fixed temperature $T$, the expansion rate is larger for a larger value of $\alpha$.  For that purpose, we define a new parameter,
\begin{eqnarray}
\label{eq18}
h(\alpha) &=& \left( \frac{H}{M_\mathrm{p}} \right)^{4 \alpha} \frac{M_\mathrm{p}^4}{8 \pi \rho} = \alpha^{4 \alpha} g_\alpha^{-1} \nonumber\\
&=& \left( \frac{\alpha}{6} \right)^{2 \alpha} \frac{ 2 
\left| 2\alpha -1 \right|
^{1-2\alpha} }{-10 \alpha^2 +8 \alpha -1 }.
\end{eqnarray}
The derivative of the $h(\alpha)$ function is given for $\alpha \neq 1/2$ by
\begin{eqnarray}
\label{eq19}
h'(\alpha) &=& \left( \frac{\alpha}{6} \right)^{2 \alpha} \frac{ 4 
\left| 2\alpha -1 \right|
^{1-2\alpha} }{-10 \alpha^2 +8 \alpha -1 } \nonumber\\
&& \times \left[\ln \frac{\alpha}{ 12 
\left| \alpha -1/2 \right| 
} +\frac{10 \left( \alpha -2/5 \right)}{-10 \alpha^2 +8 \alpha -1 } \right].~~~
\end{eqnarray}
In the allowed parameter region of Eq. (\ref{eq16}) and $0.4347 < \alpha <1/2$, the inequality $h'(\alpha) >0$ holds.  The $h(\alpha)$ and the Hubble rate are, therefore, monotonically increasing functions of $\alpha$ in the vicinity of $\alpha=1/2$.

\section{BBN calculation}\label{sec3}
The public BBN calculation code \cite{Kawano1992,Smith:1992yy} is utilized and modified in this calculation.  We updated rates of reactions related to nuclei with mass numbers $\le  10$ using the JINA REACLIB Database \cite{Cyburt2010} (the latest version taken in December 2014).  The neutron lifetime is the central value of the Particle Data Group, $880.3 \pm 1.1$~s~\cite{Agashe:2014kda}.  The baryon-to-photon ratio is $(6.037 \pm 0.077) \times 10^{-10}$ \cite{Ishida:2014wqa}, corresponding to the baryon density determined by the Planck observation of the cosmic microwave background, $\Omega_\mathrm{m} h^2 =0.02205 \pm 0.00028$ \cite{Ade:2013zuv}.

In general, BBN codes take in only two independent equations from the equations of motion and the energy conservation equation.  For example, the Kawano code \cite{Kawano1992} uses equations of the Hubble expansion rate and the time derivative of temperature, i.e., $dT/dt$.  In the present modified gravity model, the Hubble expansion rate is given by Eq. (\ref{eq17}).  The energy conservation equation, i.e., Eq. (\ref{eq7}), is, on the other hand, not changed from that in the standard BBN (SBBN) model.  We can, therefore, use the same equation for the time evolution of temperature as that in the SBBN model [Eq. (D.26) in Ref. \cite{Kawano1992}].  Then, only one modification of the Hubble rate to the code is required for BBN network calculations.

\section{Observed light element abundances}\label{sec4}
Calculated BBN results are compared to the following observational constraints on light element abundances.
The primordial $^4$He abundance is estimated with observations of metal-poor extragalactic
H II regions.  We use the latest determination of $Y_{\rm p}=0.2551\pm 0.0022$~\cite{Izotov:2014fga}.
The primordial D abundance is estimated with observations of metal-poor Lyman-$\alpha$ absorption systems in the foreground of quasistellar objects.  We use the weighted mean value of D/H$=(2.53 \pm 0.04) \times 10^{-5}$~\cite{Cooke:2013cba}.
$^3$He abundances are measured in Galactic H II regions through the $8.665$~GHz
hyperfine transition of $^3$He$^+$.  These are, however, not the primordial abundances but present values which have also been also affected by Galactic chemical evolution, including production and destruction in stars.  Nevertheless, it is very hard to reduce elemental abundances in the whole Galaxy, which have increased once by a significant factor since almost all of gas in the Galaxy needs to be incorporated into the stars multiple times and experience nuclear destruction reactions.   We then adopt only the upper limit from the abundance $^3$He/H=$(1.9\pm 0.6)\times 10^{-5}$~\cite{Bania:2002yj} in Galactic H II regions.  We note that this is neither a significant nor a direct limit on the primordial abundance and should be considered to be just a rough guide.
The primordial $^7$Li abundance is estimated with observations of Galactic metal-poor stars.  We use the abundance $\log(^7$Li/H)$=-12+(2.199\pm 0.086)$ derived in a 3D nonlocal thermal equilibrium model~\cite{Sbordone2010}.
$^6$Li abundances in Galactic metal-poor stars have also been measured.   We adopt the least-stringent $2~\sigma$ upper limit of all limits for stars reported in \cite{Lind:2013iza}, i.e., $^6$Li/H=$(0.9\pm 4.3)\times 10^{-12}$ for the G64-12 (nonlocal thermal equilibrium model with five free parameters).

\section{Result}\label{sec5}
\begin{figure}
\begin{center}
\includegraphics[width=7.5cm,clip]{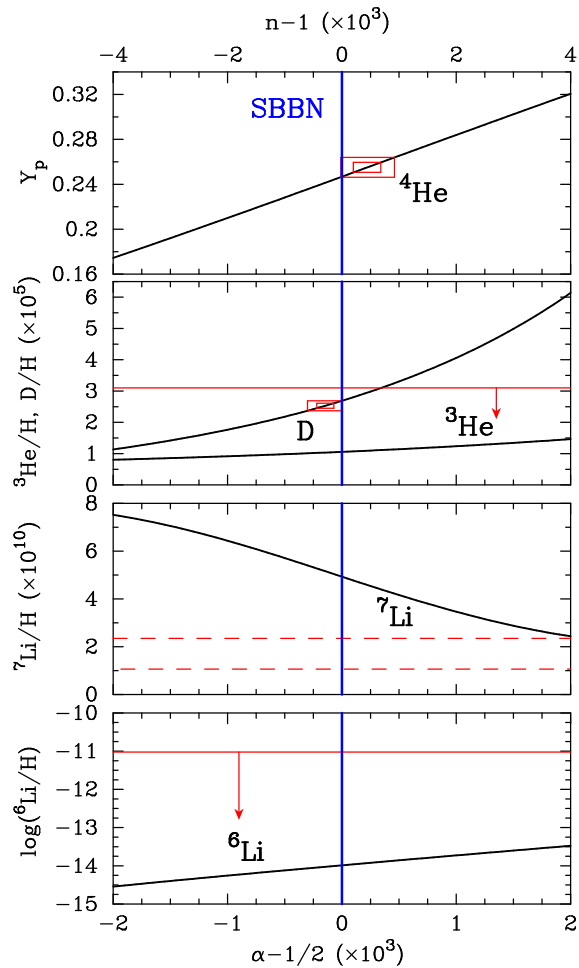}
\caption{$^4$He mass fraction $Y_\mathrm{p}$ and number abundance ratios of D, $^3$He,
 $^7$Li and $^6$Li relative to H as a function of the power-law index of the scale factor $\alpha-1/2$ or the $f(R)$ function $n-1$.  Solid curves show calculated results for the $f(R) \propto R^n$ model.  The solid smaller and larger boxes of $^4$He and D abundances correspond to the $2\sigma$ and $4\sigma$ limits, respectively, from adopted observational constraints.  The dashed box corresponds to the $2\sigma$ limits of $^7$Li abundance.  The horizontal lines with downward arrows of $^3$He and $^6$Li abundances show observational upper limits.  The result on the vertical line for $\alpha=1/2$ or $n=1$ is for the SBBN model.  Note that the $f(R)$ function for $\alpha <1/2$ was corrected in this paper so that the cosmological model connects to the present $\Lambda$CDM model.
\label{fig1}}
\end{center}
\end{figure}

Figure \ref{fig1} shows the abundances of $^4$He ($Y_{\rm p}$; mass fraction), D, $^3$He, $^7$Li, and $^6$Li (number ratio relative to H) as a function of the power-law index of the scale factor $\alpha-1/2$ or the index $n-1$ in the $f(R)$ function.  The left half of the parameter region, i.e., $\alpha <1/2$, is consistent with the present $\Lambda$CDM model when the function $f(R)$ [Eq. (\ref{eq_a4})] is adopted, while it is inconsistent when the function $f_1(R)$ [Eq. (\ref{eq_a2})] is adopted (see Sec. \ref{sec2}).


It is seen that the $2\sigma$ observational limits of $^4$He and D abundances are slightly inconsistent with the theoretical values for the SBBN, while their $4\sigma$ limits are consistent.  In this figure, however, relatively small uncertainties coming from adopted nuclear reaction rates are not taken into account.  Also, observational abundance determination may include some unknown systematic error.  In addition, the disagreements of abundances are of the order of 10 \% at most.  The slight disagreements, therefore, do not seem so important at the moment, although they could be meaningful defects in the SBBN model in the future.  In contrast to the disagreements, the $^7$Li abundance in the SBBN model is larger than observational limits by a factor of $\sim 3-4$.  The SBBN values of $^3$He and $^6$Li abundances are, on the other hand, consistent with observational upper limits.

As proven in Sec. \ref{sec2}, the cosmic expansion rate in the modified gravity model is increased when $\alpha$ increases.  In particular, the expansion rate for $\alpha>1/2$ is always larger than in the SBBN model.  All curves are understood as a result of a changed expansion rate as follows.
First, when the expansion rate is larger, the freeze-out of weak reactions occurs earlier.  The neutron abundance remaining after the freeze-out is then higher.  Second, the time interval between the freeze-out and the $^4$He synthesis is shorter because of faster cosmic expansion.  Neutron abundances are larger because of the above two reasons.  Almost all neutrons are processed to form $^4$He nuclei at the $^4$He synthesis epoch.  The $^4$He abundance is therefore larger for larger values of $\alpha$.

Because of the larger expansion rate, the reaction $^1$H($n,\gamma$)$^2$H also freezes out earlier.  The relic neutron abundance right after the $^4$He synthesis then becomes higher.  This higher neutron abundance significantly affects abundances of other light nuclei.  D is predominantly produced via $^1$H($n,\gamma$)$^2$H.  The higher neutron abundance then leads to higher D abundance.  $^3$H is produced via $^2$H($d,p$)$^3$H and destroyed via $^3$H($d,n$)$^4$He.  The enhanced D abundance leads to a higher $^3$H abundance by a higher production rate.  $^3$He is produced via $^2$H($d,n$)$^3$He and destroyed via $^3$He($n,p$)$^3$H.  The somewhat higher D abundance leads to a higher production rate, while the higher neutron abundance leads to a significantly higher destruction rate.  Eventually, the $^3$He abundance is slightly higher.  The primordial $^3$H abundance is the sum of $^3$H and $^3$He produced during the BBN.  Long after the BBN, $^3$He nuclei $\beta$-decay into $^3$H nuclei.  The final abundance of $^3$He is larger than that of $^3$H by about 2 orders of magnitude in SBBN.  Therefore, the primordial $^3$H abundance predominantly reflects the larger $^3$He abundance during BBN.

$^7$Li is produced via $^4$He($t,\gamma$)$^7$Li and destroyed via $^7$Li($p,\alpha$)$^4$He.  The $^7$Li abundance is then higher because of the higher T abundance.  $^7$Be is produced via $^4$He($^3$He$,\gamma$)$^7$Be and destroyed via $^7$Be$(n,p$)$^7$Li.  A slightly higher abundance of $^3$He and a considerably higher abundance of the neutron leads to a smaller $^7$Be abundance.  The primordial $^7$Li abundance is the sum of $^7$Li and $^7$Be produced during the BBN.  Long after the BBN, $^7$Be nuclei recombine with electrons and are transformed to $^7$Li nuclei via the electron capture process.  The abundance of $^7$Be is larger than that of $^7$Li in SBBN.  Therefore, the larger expansion rate results in a smaller primordial $^7$Li abundance.
$^6$Li is produced via $^4$He($d,\gamma$)$^6$Li and destroyed via $^6$Li$(p,\alpha$)$^3$He.  The higher D abundance leads to a higher $^6$Li abundance.

We note that trends of theoretical curves in the modified gravity model are somewhat similar to those in the model including additional components of energy density, e.g., sterile neutrinos and a primordial magnetic field, within the framework of Einstein's general relativity.  A more illustrative and detailed explanation can be found for effects of the magnetic field on all light element abundances in Ref. \cite{Kawasaki:2012va}.

When the $4\sigma$ limit of $^4$He is used, 
the constraints are derived,
\begin{eqnarray}
\label{eq_a6}
-1 \times 10^{-5} \lesssim &\left( \alpha-1/2 \right) &\lesssim 5\times 10^{-4} \nonumber \\
-2 \times 10^{-5} \lesssim &\left( n-1 \right)& \lesssim 10^{-3}.
\end{eqnarray}
When the $4\sigma$ limit of D is used, however, more stringent constraints are derived,
\begin{eqnarray}
\label{eq_a7}
-3 \times 10^{-4} \lesssim &\left( \alpha-1/2 \right)& \lesssim 2\times 10^{-6} \nonumber \\
-6 \times 10^{-4} \lesssim &\left( n-1 \right)& \lesssim 4\times 10^{-6}.
\end{eqnarray}
We find that the cases of $\alpha >1/2$ and $\alpha <1/2$ are constrained most stringently from the abundance constraints on D and $^4$He, respectively, for the following reasons.  The $4\sigma$ range of D abundance is barely consistent with the SBBN result, and the upper limit is very close to the SBBN value.  Since the D abundance increases with increasing $\alpha$ or $n$, the theoretical curve easily deviates from the observational limit for $\alpha >1/2$ (Fig. \ref{fig1}).  On the other hand, the lower limit of $^4$He abundance is very close to the SBBN value.  Since the $^4$He abundance also increases with increasing $\alpha$ or $n$, the theoretical curve easily deviates from the observational limit for $\alpha <1/2$.  

Additionally, we find that the abundances of $^3$He and $^6$Li are not sensitive to the change of $\alpha$, and that they are within the adopted limits in the whole parameter region shown in Fig. \ref{fig1}.  Although the $^7$Li abundance approaches the observed abundance level with increasing $\alpha$, it is always outside of the adopted limits in Fig. \ref{fig1}.

\begin{figure}
\begin{center}
\includegraphics[width=7.5cm,clip]{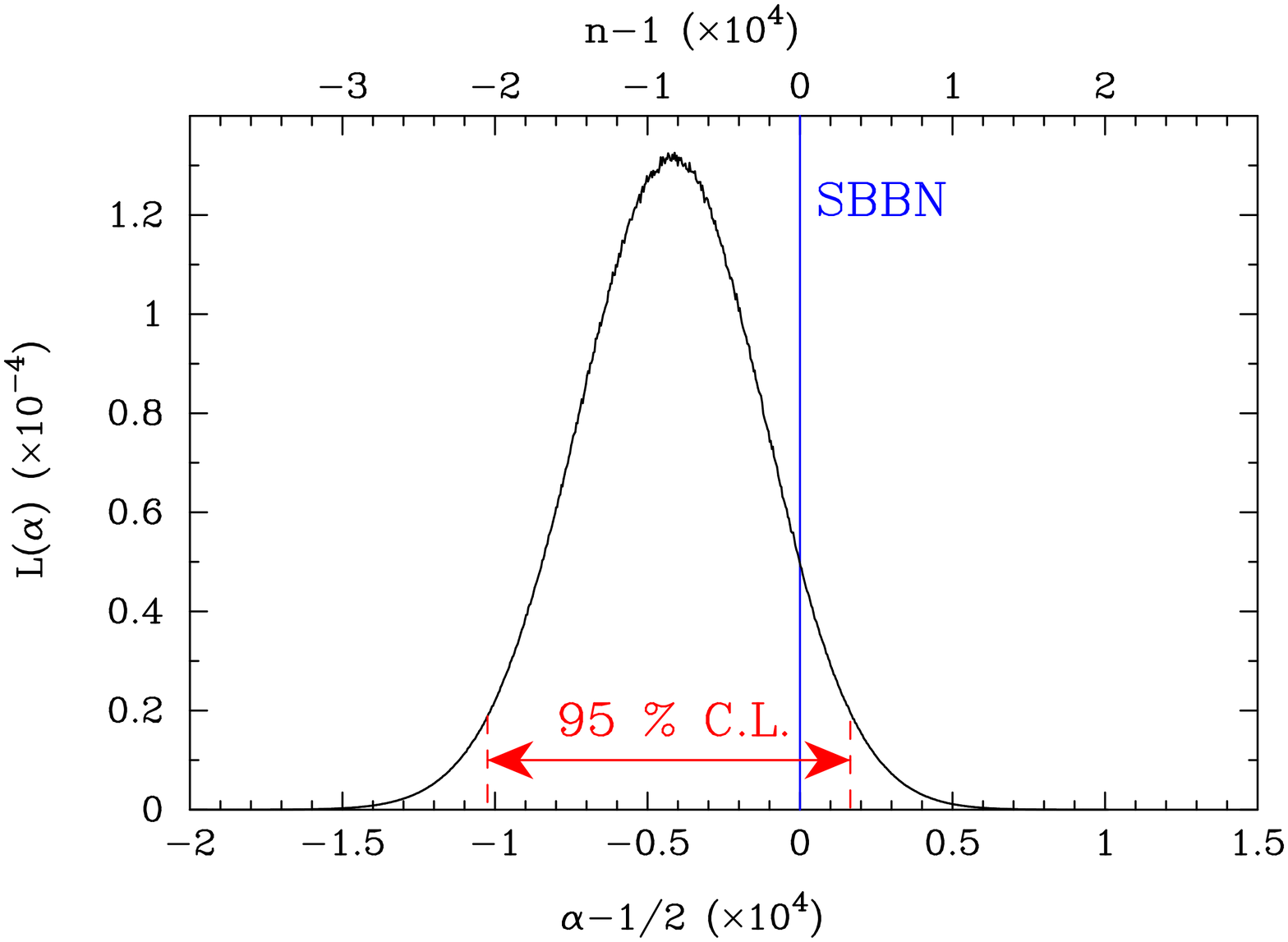}
\caption{Likelihood function as a function of the power-law index $\alpha-1/2$ or $n-1$ (solid curve).  This curve is estimated using the observational constraints on the D and $^4$He abundances.  The parameter region bounded by vertical dashed lines is the 95 \% C.L.  The vertical solid line on $\alpha=1/2$ or $n=1$ corresponds to the SBBN model.
\label{fig2}}
\end{center}
\end{figure}

Figure \ref{fig2} shows the likelihood function for the power-law index $\alpha-1/2$ or $n-1$ (solid curve).  In this estimation, the likelihood function is defined as
\begin{equation}
\label{eq_a8}
L(\alpha) =L_\mathrm{D}(\alpha) L_{^4\mathrm{He}}(\alpha),
\end{equation}
where the likelihood functions for respective nuclear abundances ($i=$D and $^4$He) are given by
\begin{equation}
\label{eq_a9}
L_i(\alpha) = \frac{1}{\sqrt{\mathstrut 2 \pi} \sigma_{i,{\rm obs}}} 
\exp \left\{ - \frac{ \left[Y_{i,{\rm th}}(\alpha) -Y_{i,{\rm obs}} \right]^2} {2 \sigma_{i,{\rm obs}}^2} \right\},
\end{equation}
where
$Y_{i,{\rm th}}(\alpha)$ is the theoretically calculated abundance of $i$, and
$Y_{i,{\rm obs}}$ and $\sigma_{i,{\rm obs}}$ are the central value and 1 $\sigma$ error, respectively, of the adopted observational abundance of $i$.  The solid curve has been normalized as $\int L(\alpha) d\alpha =1$.


From the combination of the constraints on D and $^4$He abundances, we derive a 95 \% C.L. on the parameters as
\begin{eqnarray}
\label{eq_a10}
\left( \alpha-1/2 \right) &=&\left( -0.43 \pm 0.59 \right) \times 10^{-4} \nonumber\\
\left( n-1 \right) &=&\left( -0.86 \pm 1.19 \right) \times 10^{-4}.
\end{eqnarray}
This parameter region is enclosed by vertical dashed lines.  The solid vertical line corresponds to the SBBN model described by the general relativity ($\alpha=1/2$ or $n=1$).  It is found that the deviation of the parameter $\alpha$ from $1/2$ is constrained to be less than ${\mathcal O}$($10^{-4}$).  This constraint also indicates that even in the generalized model of the $f(R)$ gravity, the case of $n=1$ is very likely based upon the comparison of theoretical and observational nuclear abundances.  If only the region of $\alpha>1/2$ is allowed, as in the case of the $f_2(R)$ function [Eq. (\ref{eq8})], the amplitude of $\alpha-1/2$ is constrained relatively strongly.

\section{Summary}\label{sec6}
In this study we revisited effects of a modified gravity on BBN.  The model is based on the $f(R) \propto R^n$ term in the action and the assumption of the scaling for the time evolution of scale factor $a(t) \propto t^\alpha$ in a homogeneous and isotropic universe.  The $f(R)$ functions were constructed so that they describe the accelerated expansion of the present Universe successfully.  We utilized a nuclear reaction network code and calculated all light element abundances in the modified gravity model.  We compared the calculations with astronomical observations of primordial elemental abundances and found that the parameters are constrained to be $\left( \alpha-1/2 \right) < {\mathcal O}(10^{-4})$ and $\left( n-1 \right) < {\mathcal O}(10^{-4})$ mainly from the limits on primordial D and $^4$He abundances.

\begin{acknowledgments}
This work was supported in part by the National Research Foundation of Korea (NRF) (Grants No. NRF-2012R1A1A2041974 and No. NRF-2014R1A2A2A05003548).  S.K. was supported by the Basic Science Research Program through the NRF funded by the Ministry
of Education (No. NRF-2014R1A1A2059080).

\end{acknowledgments}



\end{document}